\documentstyle[12pt,aasms4]{article}

\def\be{\begin{equation}}
\def\ee{\end{equation}}
\def\ba{\begin{eqnarray}}
\def\ea{\end{eqnarray}}

\def\Msolar{M_{\odot}}
\def\xibf{\mbox{\boldmath $\xi$}}
\def\etabf{\mbox{\boldmath $\eta$}}
\def\alphabf{\mbox{\boldmath $\alpha$}}
\def\xbf{{\bf x}}

\def\fun#1#2{\lower3.6pt\vbox{\baselineskip0pt\lineskip.9pt
        \ialign{$\mathsurround=0pt#1\hfill##\hfil$\crcr#2\crcr\sim\crcr}}}

\slugcomment{To be submitted to MNRAS}

\begin{document}
\null\vspace{-62pt}
\begin{flushright}
\today
\end{flushright}

\title{Caustics, Critical Curves and Cross Sections for Gravitational Lensing by Disk Galaxies}

\author{Yun Wang and Edwin L. Turner}
\affil{{\it Princeton University Observatory} \\
{\it Peyton Hall, Princeton, NJ 08544\\}
{\it email: ywang,elt@astro.princeton.edu}}

\vspace{.4in}
\centerline{\bf Abstract}
\begin{quotation}
We study strong gravitational lensing
by spiral galaxies, modeling them as infinitely thin
uniform disks embedded in singular isothermal spheres. 
We derive general properties of the critical curves and caustics
analytically.
The multiple-image cross section is a sensitive function of 
the inclination angle of the disk relative to the observer. 
We compute the inclination averaged cross section
for several sets of lensing parameters.
For realistic disk mass and size parameters, we find that the
cross section for multiple imaging is increased by only a modest
factor and {\it no} dramatic increase in the optical depth for
strong lensing of QSOs would be expected.
However, the cross section for high magnifications is 
significantly increased due to the inclusion of a disk, especially
for nearly edge-on configurations;
due to the strong observational selection effects favoring
high magnifications, there might be significant
consequences for lensing statistics.

\end{quotation}


\section{Introduction}

Early theoretical studies of gravitational lensing of distant quasars
by foreground galaxies approximated their mass distributions as spherical
and usually also as singular and isothermal (\cite{GG74}; \cite{Turner80};
Turner, Ostriker \& Gott 1984).  Later studies explored the
relaxation of these simplifying approximations and, in particular,
considered elliptical mass distributions and potentials ({\it e.g.},
\cite{BK87}, \cite{KB87}, \cite{BN92}),
but it has been shown that the standard spherical approximations are adequate
for the purpose of many statistical calculations
({\it e.g.}, \cite{FT91}, \cite{MR93}, \cite{Koch93}).
However, with rare exception (\cite{OV90}), the gravitational
lensing effects of the defining component of spiral galaxies, their very
thin disks, has been ignored in discussions of lensing statistics.
There are two reasons.  First, {\it within the limitations of the
spherical galaxy approximation}, it was early shown (Turner et al. 1984)
and later repeatedly confirmed by more detailed studies 
(\cite{FT91}, \cite{Koch91}) that spiral galaxies contribute
only a small fraction, of order 10\%, of the total lensing cross section
of the known galaxy population, with ellipticals and S0's being the
dominant contributors.  Second, the disk
stability arguments originated by Ostriker \& Peebles (1973) and the
many other lines of evidence which indicate the presence of massive
dark halos suggest that even spiral galaxy mass distributions are dominated
by a roughly spherical component.

While these considerations remain valid, there now seem to be some reasons
to examine the lensing effects of thin disk mass distributions more
carefully.  For one, samples of galaxy lensed quasars are becoming
rapidly larger and better controlled.  Thus, smaller and more subtle
features of lens statistics are becoming observationally accessible and
therefore theoretically interesting.  For another, in the usual (and very
accurate) {\it thin lens} approximation, it is not the three dimensional
mass distribution of the lensing object that matters but rather the
projected two dimensional one, particularly those regions which exceed
the lensing critical surface density (\cite{TOG84}).  From this
two dimensional point of view, a very thin disk may be much more important
when seen nearly edge-on than it is in three dimensions.  Finally,
although the situation is far from unambiguous, observations suggest
that lensing galaxies may have more dust (\cite{Law95},
Malhotra, Rhoads \& Turner 1997) and larger quadrupole mements
(Keeton, Kochanek \& Seljak 1996) than would have been naively expected if early type
galaxies entirely dominated the quasar-galaxy lensing rate.

Therefore, in this paper we present a preliminary exploration of the
lensing properties of spiral galaxies, which we approximate as finite,
uniform surface density disks of zero thickness embedded in singular
isothermal spheres (SIS).  We focus on the general caustic and critical
line properties of such lensing objects and also study the lensing
cross section enhancement over the SIS only approximation.  We
explicitly carry the important variable of disk inclination angle
throughout the calculations.  In the present paper, however, we do
not present the fully detailed calculations required for direct
comparison to observations; in particular, we do not here consider
line-of-sight integrations, integration over realistic distributions
of galaxy properties nor the effects of amplification (magnification)
bias on observed samples.  Rather, we present only an initial
qualitative and semi-quantitative investigation of lensing by disk
galaxy mass distributions.

The basic equations and model are presented in section 2.  Section 3 discusses
our technique for finding caustics, critical curves and multiple
imaging configurations.  Section 4 presents the general lensing
properties of our model of disk galaxies in the language of caustics
and critical curves.  Section 5 considers the effects of inclined disks
on lensing cross sections and the implied 
modification of quasar
lensing rates.  Finally, section 6 contains a summary of our results
and a discussion of recent related work.
We follow the conventions and notation of Schneider,
Ehlers \& Falco (1992).

\section{Light deflection due to a spiral galaxy}

We model spiral galaxies as uniform and infinitely thin disks embedded
in singular isothermal spheres.
Let us choose a length scale appropriate for a singular isothermal 
sphere (SIS), 
\be
\xi_0 = 4 \pi \left(\frac{v_{SIS}}{c} \right)^2 \frac{D_d D_{ds}}{D_s},
\ee
where $v_{SIS}$ is the line-of-sight velocity dispersion of the SIS.
Let $\xibf$ and $\etabf$ be the physical position vectors
of the image (in the lens plane) and the source (in the source plane)
respectively, then the dimensionless image and source
positions are
\be
{\bf x}=\frac{\xibf}{\xi_0}; \hskip 1cm 
{\bf y}= \frac{\etabf}{\eta_0},
\ee
where $\eta_0= \xi_0 D_s/D_d$,
$D_s$ and $D_d$ are our distances to the source and lens respectively. 
The lens equation becomes
\be
{\bf y}= {\bf x}- \alphabf({\bf x}),
\ee
The scaled deflection angle 
\be
\label{eq:alpha}
\alphabf({\bf x})= \frac{1}{\pi} \int{\rm d}^2x'\, \kappa({\bf x}')\,
\frac{ {\bf x}- {\bf x}'}{|{\bf x}- {\bf x}'|^2},
\ee
where $\kappa({\bf x})= \Sigma(\xi_0{\bf x})/\Sigma_{crit}$, with
$\Sigma_{crit}= c^2 D_s/(4\pi G D_d D_{ds})$. $D_{ds}$ is the
distance between the lens and the source.
We use affine distances, $D_d=cH_0^{-1}\lambda(z_d)$, 
$D_s= cH_0^{-1}\lambda(z_s)$, $D_{ds}= cH_0^{-1}\lambda(z_d,z_s)=
cH_0^{-1} (1+z_d)[\lambda(z_s)-\lambda(z_d)]$,
where
\be
\lambda(z)= \int_0^z {\rm d}w\, \frac{1}{(1+w)^2
\sqrt{\Omega_0 (1+w)^3 +\Omega_{\Lambda} } }.
\ee
For $\Omega_0=1$, $\lambda(z)=(2/5)\, \left[1- (1+z)^{-5/2}\right]$.

For a SIS, its dimensionless surface density and scaled deflection
angle are (\cite{book})
\be
\kappa^{SIS}({\bf x})= \frac{1}{2 |{\bf x}|}; \hskip 1cm
\alphabf^{SIS}({\bf x})= \frac{\bf x}{|{\bf x}|}.
\ee

The projection of a circular disk of radius $r_{disk}$ in the plane 
perpendicular to the line of sight is an ellipse with semi-major axes 
$r_{disk}$ and $r_{disk}\,\cos\theta$, where $\theta$ is the angle 
between the normal vector of the inclined disk plane and the line of sight.
Note that $0\leq \theta \leq \pi/2$;
$\theta=0$ is the ``face-on'' disk case; $\theta=\pi/2$ is the
``edge-on'' disk case.

For a uniform disk, its projected surface density is 
$\Sigma_{disk} =M_{disk}/(\pi r_{disk}^2 \cos\theta)$; its projected 
dimensionless surface density is
\ba
\kappa^{disk} &= &\frac{\Sigma_{disk}}{\Sigma_{crit}}
= \frac{1}{r_{disk}^2 \cos\theta}\, \frac{4GM_{disk}\, D_d D_{ds}}{c^2 D_s},
\nonumber \\
&=& \frac{0.19}{\cos\theta} \, \left( \frac{10\,\mbox{kpc}}{r_{disk}}
\right)^2\, \left(\frac{M_{disk}}{10^{11}\,\Msolar}\right)
\,\left( \frac{D_d}{1000\, \mbox{Mpc}}\right)\,
\frac{D_{ds}}{D_s}.
\ea
For simplicity in notation, we define
\be
q = \kappa^{disk} \cos\theta.
\ee
$q$ is the {\it face-on} surface mass density of the disk.
Note that $q<1$ for realistic choices of the parameters.

The dimensionless radius of the disk is
\be
R = \frac{r_{disk}}{\xi_0}= \frac{10}{\pi} \left(\frac{r_{disk}}{10
\,\mbox{kpc}}\right)\, \left( \frac{150\,\mbox{km/s}}{v_{SIS}}
\right)^2\, \left( \frac{1000\, \mbox{Mpc}}{D_d}\right)\,
\frac{D_s}{D_{ds}}.
\ee

Let us choose the ${\bf x}$ coordinates such that the
contour of the projected disk is given by
\be
\left(\frac{x_1}{R}\right)^2+ \left(\frac{x_2}{R\cos\theta}\right)^2=1.
\ee
Then the total scaled deflection angle 
$\alphabf=\alphabf^{SIS}+\alphabf^{disk}$,
with $\alphabf^{disk}$ given by
\ba
\label{eq:alphadisk}
&&\alpha_1^{disk}= \frac{R\,q}{2\pi}
\int_{-1}^{1}{\rm d}\omega \, \ln\left[
\frac{ ( \sqrt{1-\omega^2}+ x_1/R)^2 + (\omega\cos\theta- x_2/R)^2 }
{ ( \sqrt{1-\omega^2}- x_1/R)^2 + (\omega\cos\theta- x_2/R)^2 }
\right], \nonumber \\
&&
\alpha_2^{disk}= \frac{R\,q}{2\pi\cos\theta}
\int_{-1}^{1}{\rm d}\omega \, \ln\left[
\frac{ ( \omega- x_1/R)^2 + ( \sqrt{1-\omega^2}\,\cos\theta+ x_2/R)^2 }
{ ( \omega- x_1/R)^2 + ( \sqrt{1-\omega^2}\,\cos\theta- x_2/R)^2 }
\right].
\ea
Note that 
\be 
\alpha_2^{disk}(\cos\theta \rightarrow 0)= \frac{2 x_2 q}{\pi}
\int_{-1}^{1}{\rm d}\omega \,
\frac{ \sqrt{1-\omega^2} }
{ ( \omega- x_1/R)^2 + ( \sqrt{1-\omega^2}\,\cos\theta- x_2/R)^2 }.
\ee

To study image multiplicity, we will also need the derivatives 
of the deflection angles:
\ba
\label{eq:dalpha/dx}
&&\frac{\partial\alpha_1}{\partial x_1}= \frac{2 q}{\pi}
\int_{-1}^1{\rm d}\omega\, \frac{\sqrt{1-\omega^2}}{f_1(\omega)}\,
\left[ 1-\omega^2 - \left(\frac{x_1}{R} \right)^2 +
\left(\frac{x_2}{R}-\omega\cos\theta \right)^2 \right]
+\frac{x_2^2}{x^3}, \nonumber\\
&&\frac{\partial\alpha_1}{\partial x_2}= -\frac{4 q}{\pi}
\left(\frac{x_1}{R}\right)
\int_{-1}^1{\rm d}\omega\, \frac{\sqrt{1-\omega^2}}{f_1(\omega)}\,
\left(\frac{x_2}{R}-\omega\cos\theta \right)
-\frac{x_1 x_2}{x^3}, \\
&&\frac{\partial\alpha_2}{\partial x_1}= -\frac{4 q}{\pi}
\left(\frac{x_2}{R}\right)
\int_{-1}^1{\rm d}\omega\, \frac{\sqrt{1-\omega^2}}{f_2(\omega)}\,
\left(\frac{x_1}{R}-\omega \right)
-\frac{x_1 x_2}{x^3}, \nonumber\\
&&\frac{\partial\alpha_2}{\partial x_2}= \frac{2 q}{\pi}
\int_{-1}^1{\rm d}\omega\, \frac{\sqrt{1-\omega^2}}{f_2(\omega)}\,
\left[ \left(\frac{x_1}{R}-\omega \right)^2 -
\left(\frac{x_2}{R}\right)^2 + (1-\omega^2)\cos^2\theta \right]
+\frac{x_1^2}{x^3}.\nonumber
\ea
We have defined 
\ba
\label{eq:D1,D2}
&& f_1(\omega) \equiv \left[ \left(\frac{x_1}{R}+ \sqrt{1-\omega^2}\right)^2
+\left(\frac{x_2}{R}-\omega\cos\theta\right)^2\right]\,
\left[ \left(\frac{x_1}{R} - \sqrt{1-\omega^2}\right)^2+
\left(\frac{x_2}{R}-\omega\cos\theta\right)^2\right],
\nonumber \\
&& f_2(\omega) \equiv \left[ \left(\frac{x_1}{R}- \omega \right)^2
+\left(\frac{x_2}{R}+ \sqrt{1-\omega^2}\,\cos\theta\right)^2\right]\,
\left[ \left(\frac{x_1}{R}- \omega\right)^2
+\left(\frac{x_2}{R}- \sqrt{1-\omega^2}\,\cos\theta\right)^2\right].
\nonumber\\
&&
\ea

\section{Condition for multiple images}

The magnification factor of the source is given by 
$\mu({\bf x}) = 1/\det\,A({\bf x})$, where
the Jacobian matrix $A({\bf x})$ is defined as
\ba
\label{eq:detA}
&&A({\bf x})= \frac{\partial {\bf y}}
{\partial {\bf x}}, \hskip 1cm
A_{ij}= \frac{\partial y_i} {\partial x_j};\\
&&\det A= \left(1- \frac{\partial\alpha_1}{\partial x_1}\right)\,
\left(1- \frac{\partial\alpha_2}{\partial x_2} \right)
-\frac{\partial\alpha_1}{\partial x_2}\cdot 
\frac{\partial\alpha_2}{\partial x_1}. \nonumber
\ea

It has been shown that an isolated transparent lens can produce
multiple images if, and only if, there is a point $\bf x$ with
$\det\,A({\bf x})<0$. If at ${\bf x}_0$, $\det\,A({\bf x}_0)<0$
(negative parity),
a source at ${\bf y}_0={\bf y}({\bf x}_0)$ has at least two
additional images of positive parity. The multiple-image cross-section
is simply the bounded area in the source plane in which for
each source position ${\bf y}$, there exists an image position ${\bf x}$
where $\det\,A({\bf x})<0$.

Formally, critical curves are given by $\det A(\xbf)=0$,
the corresponding source positions are the caustics.
However, the cross-section for multiple images is not always bounded
by a caustic. In the case of SIS only, $\det\,A=1-1/x$,
where $x = \sqrt{x_1^2+ x_2^2}$. $\det\,A<0$ gives $ 0\leq x < 1$.
The lens equation gives us $y=|x-1|$. Hence $y \leq 1$ for multiple
images; the multiple image cross-section is bounded by $y=1$, which corresponds
to $x=0$ (where $\det\,A=-\infty$), while the only critical curve is
at $x=1$ corresponding to $y=0$.

Next let us consider the case of SIS plus 
a face-on disk, $\theta=0$, $\kappa^{disk}=q$. It's
straightforward to integrate Eqs.(\ref{eq:alphadisk}) to find (\cite{book})
\be
\alphabf = \left\{ \begin{array} {ll}
\xbf/x + \kappa^{disk} \, \xbf, \hskip 1cm x<R; \\
\xbf/x + \kappa^{disk} R^2 \, \xbf/ x^2, \hskip 1cm x>R. \end{array}
\right.
\ee

Hence we have
\be
\det A=\left\{ \begin{array}{ll}
\left(1-\kappa^{disk}\right)\,\left(1-\kappa^{disk} - 1/x\right), 
\hskip 1cm x<R; \\
\left(1+ \kappa^{disk} R^2/x^2\right)\,
\left(1-1/x - \kappa^{disk} R^2/x^2 \right),
\hskip 1cm x>R. \end {array} \right.
\ee
It is easy to see that $\det A$ is {\it discontinuous} at $x=R$.

For $\kappa^{disk}<1$, $\det A$ does {\it not} change sign at $x=R$. 
If $R>1/(1-\kappa^{disk})$, the only critical curve is given by
$x=1/(1-\kappa^{disk}) <R$, and it maps to $y=0$;
if $R<1/(1-\kappa^{disk})$, the only critical curve is given by
$x=\left(1+ \sqrt{1+ 4\kappa^{disk} R^2} \right)/2 >R$, 
and it also maps to $y=0$.
The critical curves therefore have no relevance to image multiplicity.
On the other hand, $\det A(x)<0$ leads to $y \leq 1$, with $y=1$
given by $x=0$, just as in the SIS only case.
SIS with a face-on disk of $\kappa^{disk}<1$ has the same multiple-image 
cross section as SIS only.

If $\kappa^{disk}>1$, the true critical curve is given by 
$x=(1+ \sqrt{1+ 4\kappa^{disk} R^2} )/2 >R$, which again maps to $y=0$.
However, $\det A(x<R)>0$, $\det A(x \rightarrow R^{+})<0$,
hence $\det A$ changes sign at $x=R$, which is effectively a 
``critical curve''; the corresponding ``caustic''
is given by $y=1+R(\kappa^{disk}-1) >1$. 
SIS with a face-on disk of $\kappa^{disk}>1$ has a larger multiple-image 
cross section than SIS only.

\section{Image multiplicity for SIS plus inclined disk}

For SIS plus uniform disk, $\alphabf^{disk}(x=0)=0$, hence $x=0$ corresponds
to source position $y=1$, just as in the SIS only case. When 
$\det\,A(x\rightarrow 0)<0$, $y\leq 1$ always gives multiple images;
the multiple-image cross-section increases relative to the SIS only case
if any caustic curves lie outside the $y=1$ circle.

The value of $\det A$ at $x=0$ is a useful indicator
of the general properties of the critical curves. 
For uniform disk plus SIS, we find [using Eqs.(\ref{eq:dalpha/dx}) and
(\ref{eq:detA})]
\be
\label{eq:detA(x=0)}
\det A (x\rightarrow 0) = 1-\frac{2q}{\cos\theta}+ \frac{1}{\cos\theta}
\left(\frac{2q}{1+\cos\theta}\right)^2
 + \frac{C_1(\theta)\, x_1^2}{x^3}+ \frac{C_2(\theta)\, x_2^2}{x^3},
\ee
where
\be
C_1(\theta)=\frac{2q}{1+\cos\theta}-1, \hskip 1cm
C_2(\theta) = \frac{2q}{\cos\theta(1+\cos\theta)}-1.
\ee
Let us define
\be
S \equiv C_1 x_1^2+ C_2 x_2^2.
\ee
$S$ has the same sign as $\det A (x\rightarrow 0)$.
We have two critical angles:
\ba
\label{eq:thetacrit}
&& C_1(\theta)=0: \hskip 1cm \theta=\theta_1= \arccos(2q-1);\nonumber\\
&& C_2(\theta)=0: \hskip 1cm \theta=\theta_2= 
\arccos\left( \frac{\sqrt{1+8q}-1}{2} \right).
\ea
Note that $\theta_1$ is not defined for $q<1/2$.
For $q>1/2$, $\theta_1>\theta_2$.

Note that $\det A$ is {\it discontinuous} for $(x_1^c,x_2^c)$ on the ellipse
$(x_1^c/R)^2+ (x_2^c/R\cos\theta)^2=1$ [see Eqs.(\ref{eq:D1,D2})],
i.e., on the edge of the disk, just as in the face-on disk case.
If $\det A(x\rightarrow 0) = +\infty$, $\det A $ changes sign at
$(x_1^c,x_2^c)$, which describes effectively a ``critical curve'',
and the corresponding source positions give us a ``caustic''.

First we consider $q<1/2$. Here we always have $C_1<0$.
We have two different cases: (1) $\theta<\theta_2$, $C_2<0$, $S<0$,
$\det A(x\rightarrow 0) = -\infty$, we have only one true critical curve 
and it encloses $x=0$;
(2) $\theta>\theta_2$, $C_2>0$, $S>0$ [$\det A(x\rightarrow 0) = +\infty$] 
if $|x_2|> \sqrt{|C_1|/C_2}\, |x_1|$, $S<0$ 
[$\det A(x\rightarrow 0) = -\infty$] 
if $|x_2|< \sqrt{|C_1|/C_2}\, |x_1|$, 
here we have one true critical curve enclosing $x=0$, plus an additional
``critical curve'' which is hourglass shaped and passes through $x=0$,
and is closed off by $(x_1^c,x_2^c)$ defined above.
Figs.1-3 show the critical curves (a) and caustics (b) for $q=0.2$
($\theta_2=72.168^{\circ}$), for $\theta=70^{\circ}$,
75$^{\circ}$, and 85$^{\circ}$ respectively. The unit radius circle in 
(a) indicates the critical curve for the SIS only case (its corresponding
caustic shrinks to the point $y=0$ in the source plane); the
unit radius circle in (b) indicates the multiple-image cross-section
for the SIS only case.

For $q>1/2$, we have three different cases: 
(1) $\theta<\theta_2$, $C_1<0$, $C_2<0$, $S<0$,
$\det A(x\rightarrow 0) = -\infty$, we have only one true critical curve 
and it encloses $x=0$;
(2) $\theta_2<\theta< \theta_1$, $C_1<0$, $C_2>0$, 
$S>0$ [$\det A(x\rightarrow 0) = +\infty$] 
if $|x_2|> \sqrt{|C_1|/C_2}\, |x_1|$, 
$S<0$ [$\det A(x\rightarrow 0) = -\infty$] 
if $|x_2|< \sqrt{|C_1|/C_2}\, |x_1|$, 
here we have one true critical curve enclosing $x=0$, plus an additional
``critical curve'' which is hourglass shaped and passes through $x=0$,
and is closed off by $(x_1^c,x_2^c)$ defined above;
(3) $\theta>\theta_1$, $C_1>0$, $C_2>0$, $S>0$,
$\det A(x\rightarrow 0) = +\infty$, we have two critical curves 
enclosing $x=0$, one of them is the true critical curve, the other is
given by $(x_1^c,x_2^c)$ defined above.
Figs.4-6 show the critical curves (a) and caustics (b) for $q=0.6$
($\theta_1=78.463^{\circ}$, $\theta_2= 45.238^{\circ}$), 
for $\theta=45^{\circ}$, 75$^{\circ}$, and 85$^{\circ}$ respectively. 
The unit radius circles in the
figures are the same as already indicated.

Because $\det A(\xbf)$ is modified from Eq.(\ref{eq:detA(x=0)})
as we move away from $x=0$, the hourglass shaped ``critical curves''
are given by the intersection of $|x_2| \sim \sqrt{|C_1|/C_2}\, |x_1|$ and
the ellipse $(x_1^c/R)^2+ (x_2^c/R\cos\theta)^2=1$; this explains
the curved sides of the hourglass shape.
Due to numerical noise, some discrete points from the ($x_1^c, x_2^c$)
ellipse and their corresponding source positions appear in  
Figs.1-6 as discrete points apart from the critical curves and caustics.

It is interesting to note that our study of the face-on disk case
(see the previous section)
suggests that the critical angle for the disk to contribute to
image multiplicity is given by $\kappa^{disk}=q/\cos\theta=1$,
i.e., $\theta=\theta_0 \equiv \arccos(q)$. For $q<1/2$, only 
$\theta_2$ is defined, and $\theta_0>\theta_2$;
for $q>1/2$, $\theta_1> \theta_0 >\theta_2$.
This indicates that an inclined disk is more efficient than a face-on
disk with the same surface mass density.

\section{Increased multiple-image cross section due to the inclined disk}

When there is only {\it one} critical curve (the true critical curve)
in the ${\bf x}$ plane, the multiple image cross-section
is given by the area enclosed by $y=1$ plus the areas enclosed by the caustic
which lies {\it outside} the $y=1$ circle, which are two small sharp-angular 
areas lying along the $y_1$ axis; 
the caustic is completely inside the $y=1$ circle for
sufficiently small $\theta$.
When we have {\it two} branches of
critical curves (one of them is a true critical curve, the other
corresponds to discontinuity and change of sign in $\det A$)
in the ${\bf x}$ plane, the multiple image cross-section
is given by the area enclosed by $y=1$ plus the areas enclosed by the caustics
which lie {\it outside} the $y=1$ circle, which now consist of two angular
areas lying along the $y_1$ axis 
(corresponding to the true critical curve, at largest possible $x$), 
plus two round areas lying along the $y_2$ axis
(corresponding to the ``critical curve'' due to discontinuity
and change of sign in $\det A$).

Assuming uniform distribution of the inclination angle $\theta$ in solid angle,
the average multiple-image cross section is
\be
\overline{\sigma}_{\theta}= \frac{ \int {\rm d}\theta \, \sin\theta \, 
\sigma(\theta)}
{ \int {\rm d}\theta \, \sin\theta },
\ee
where $\sigma(\theta)$ is multiple-image cross section for inclination 
angle $\theta$.

Figs.7-9 show the ratio of the multiple-image cross-sections for
SIS plus inclined uniform disk and for SIS only, $\sigma(\theta)/\sigma^{SIS}$,
as a function of the inclination angle $\theta$, for three groups (nine sets) of
choices of ($q,R)$. Table 1 lists the parameters and the corresponding
average enhancement factor in multiple-image cross section
$\overline{\sigma}_{\theta}/\sigma^{SIS}$.

\begin{table}[h]
\caption{Average enhancement factor in multiple-image cross section}
\begin{center}
\begin{tabular}{|c||p{1in}|p{1in}|p{1in}|}
\hline
  & $R=1.5$ & $R=3$ & $R=6$ \\ \hline\hline
q=0.2 &  1.086  &  1.160 &  2.041 \\ \hline
q=0.4 &  1.296  &  1.547 &  2.742 \\ \hline
q=0.6 &  1.718  &  2.493 &  4.509 \\ \hline
\end{tabular}
\end{center}
\end{table}

Let us write
\be
q\,R= \frac{1.9}{\pi}\, \left( \frac{10\,\mbox{kpc}}{r_{disk}}\right)\,
\left( \frac{ M_{disk}}{ 10^{11} \Msolar} \right)\,
\left( \frac{150\,\mbox{km/s}}{v_{SIS}} \right)^2.
\ee
For a model spiral galaxy with given ($v_{SIS}, M_{disk}, 
r_{disk}$), $q\,R$ is constant.
Inspection of Table 1 shows that the modification of the galaxy's inclination
averaged cross section is quite small, less than say 50\% (there are other
uncertainties in lens statistics calculations {\it at least} this large), 
if $q\,R$ is less than about unity.  From equation (23) we then see that
significant modification of the multiple image lensing cross section will
only occur for objects with uncharacteristically small and massive disks and/or
those with minimal spherical (halo) components.  Few, if any, real galaxies
may satisfy such conditions.

The differential probability of a strong-lensing (multiple-images)
event is (\cite{TOG84})
\be
{\rm d}\tau = n_L(z_d)\, \overline{\sigma}_{\theta}(z_d|z_s)\,
\frac{c\, {\rm d}t}{ {\rm d} z_d}\, {\rm d} z_d,
\ee
where $n_L(z_d)$ is the number density of lenses at lens redshift $z_d$,
$\overline{\sigma}_{\theta}(z_d|z_s)$ is the inclination-angle averaged 
cross-section for multiple images given source redshift $z_s$.
From the above discussion, we would expect the total contribution of spiral
galaxies to the QSO strong lensing optical depths to increase only slightly
due to the effects of their thin disks.  The optical depths would then continue
to be dominated by early type galaxies.

However, the inclusion of an inclined disk breaks the circular
symmetry due to the SIS, the true critical curve [where $\det A(x)=0$]
no longer maps to a point ($y=0$) as in the SIS only case,
but maps to a caustic which encloses an area comparable to
the SIS only multiple-image cross-section for $\theta>\theta_2$
[see Eq.(\ref{eq:thetacrit})].
Since the magnification is infinite (for a point source) on the caustic,
and decreases smoothly away from it,
the cross section for high magnifications is significantly increased
due to the inclusion of a disk.
Quantitative investigation of this effect will require extensive numerical
calculations which are outside the scope of this paper, but 
this effect may well be important.  
Observational selection effects favoring inclusion of high
amplification (magnification) lensing configurations in flux limited
samples can lead to major distortions of intrinsic distributions in real
samples; see Ostriker \& Vietri (1990), for example.

\section{Discussion and conclusion}

We have examined strong (multiple image) gravitational lensing
by spiral galaxies, modeled as infinitely thin
uniform disks embedded in singular isothermal spheres. 
We have derived general properties of the critical curves and caustics
analytically.
The multiple-image cross section is a sensitive function of 
the inclination angle of the disk relative to the observer. 
We have therefore computed the inclination averaged cross section
for several sets of lensing parameters.

We find that
the optical depth for multiply imaged QSOs should only
increase by a factor of a few at most and by less
than 50\% in nearly all realistic cases; the inclusion of
a disk is therefore expected to have {\it no} significant effect 
on the contribution
of spiral galaxies to the total optical depth for
multiply imaged QSOs.
On the other hand, the cross section for high magnifications is 
significantly increased due to the inclusion of a disk.
The increase in the number of lensed QSOs with
high magnifications could have substantial effects on observed lens samples.

While completing this work, we became aware of recent preprints
by Loeb (1997) and by
Maller, Flores, and Primack (1997). 
The former is primarily concerned with the possible connectin between
lensing by spiral galaxies and high column density HI absorption systems
seen in QSO spectra but also briefly mentions the potential role of disk
inclination in spiral galaxy lensing statistics.
The subject of
our paper overlaps partially with that of Maller {\it et al.}, 
and some of our results
are qualitatively similar;
however, our paper is mainly analytical (which provides insight into the
mathematics of the critical curves and caustics) while theirs is numerical.


We thank A. Loeb for useful discussions and gratefully acknowledge 
support from NSF grant AST94-19400.




\clearpage


\setcounter{figure}{0}

\plottwo{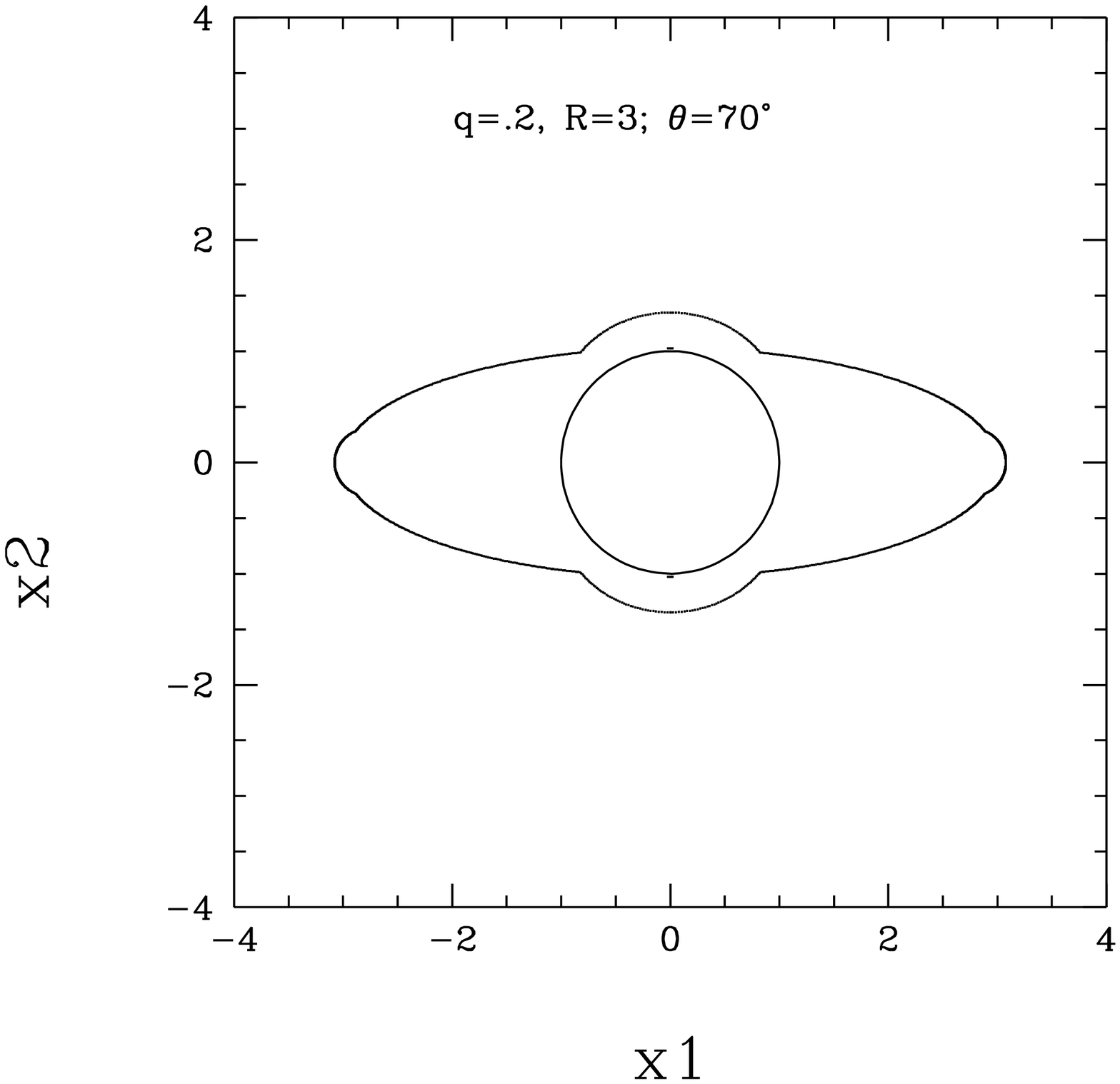}{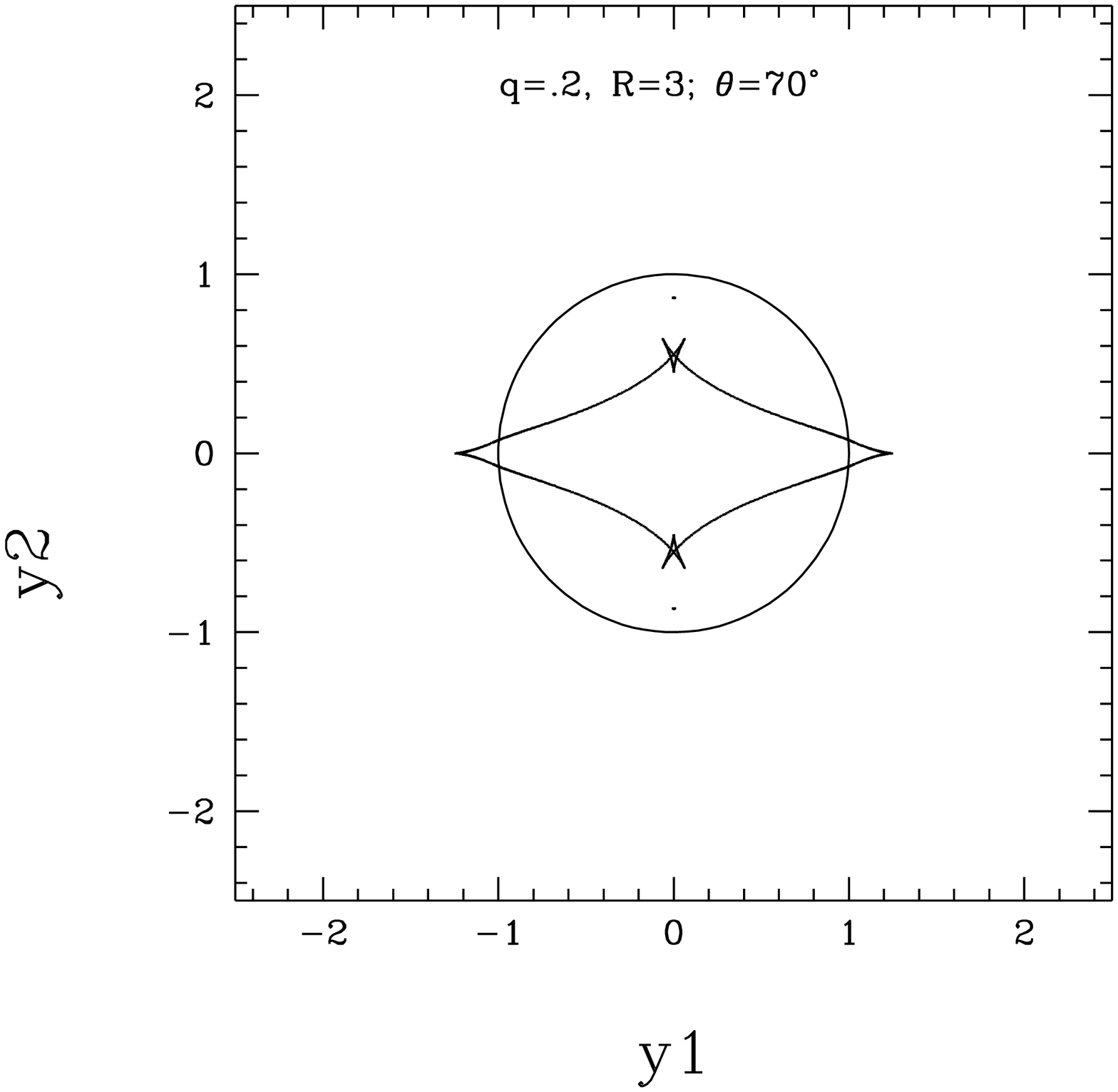}
\figcaption{$q=0.2$, $R=3$, 
$\theta=70^{\circ}<\theta_2=72.168^{\circ}$.
(a) Critical curve. The unit radius circle indicates the critical curve 
for the SIS only case (its maps to $y=0$ in the source plane).
(b) Caustic. The unit radius circle indicates the multiple-image 
cross-section for the SIS only case.}

\plottwo{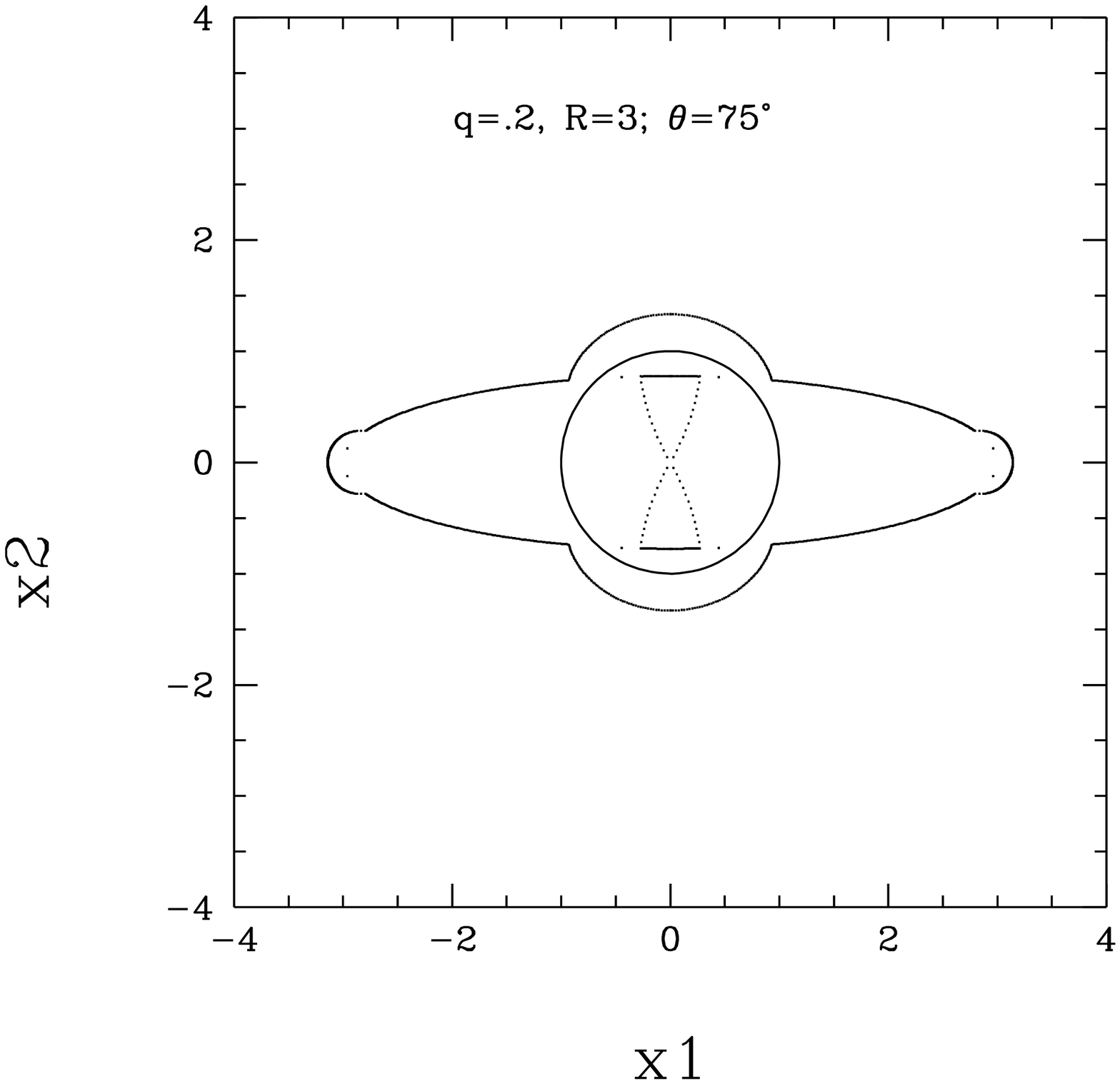}{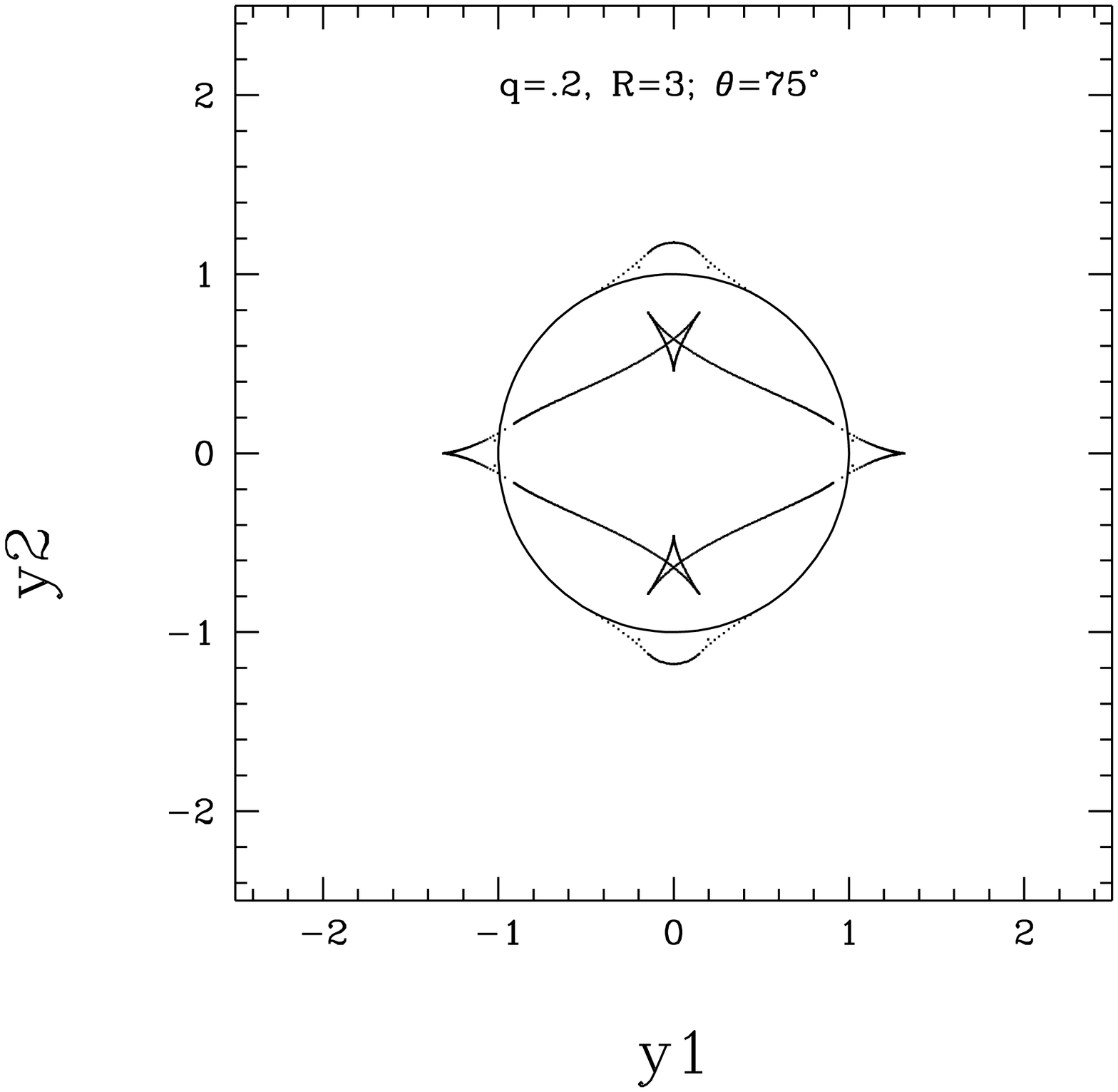}
\figcaption{$q=0.2$, $R=3$, 
$\theta=75^{\circ}> \theta_2=72.168^{\circ}$. 
The unit radius circles are the same as in Fig.1.
(a) Critical curve. (b) Caustic. }

\plottwo{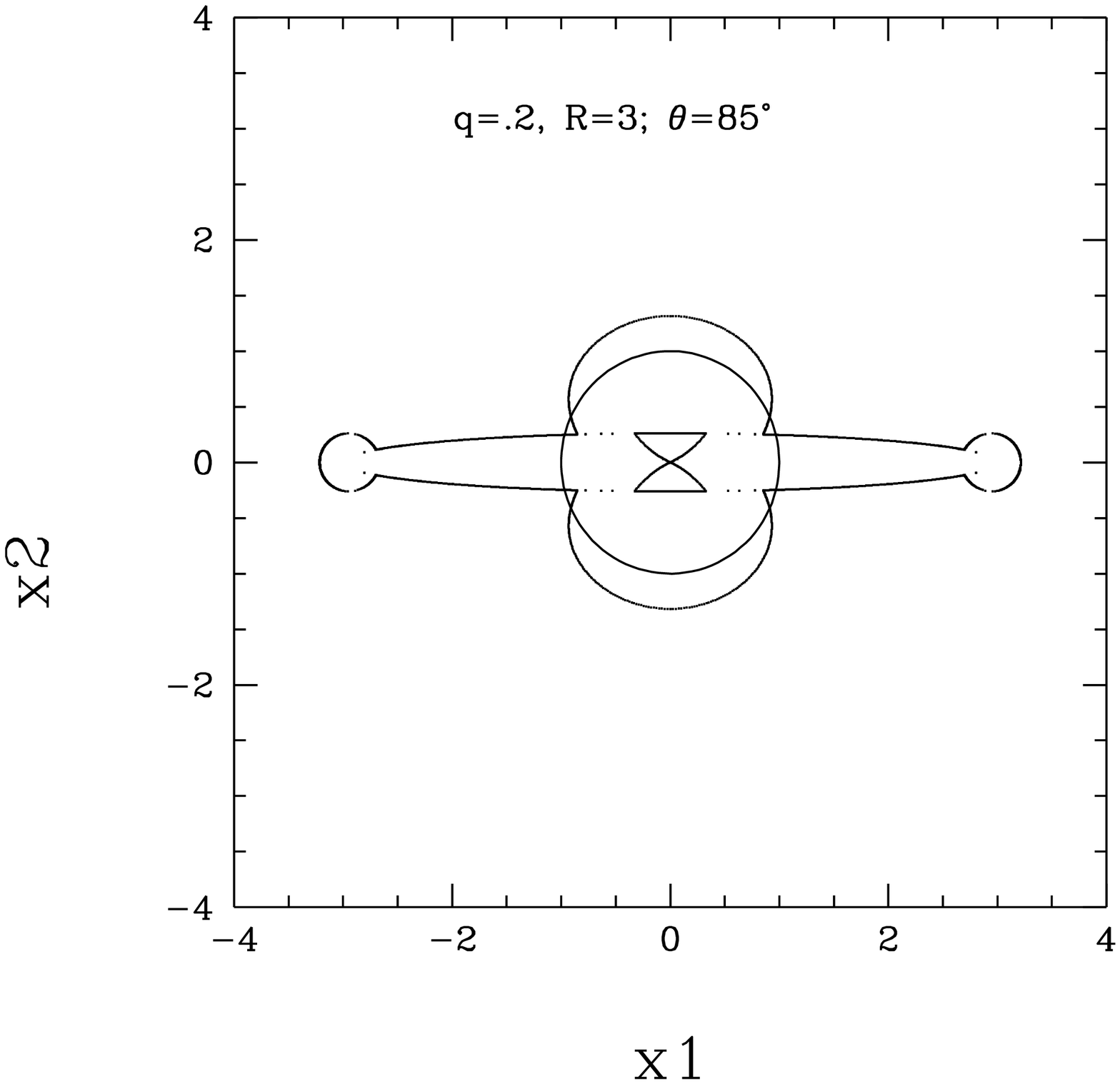}{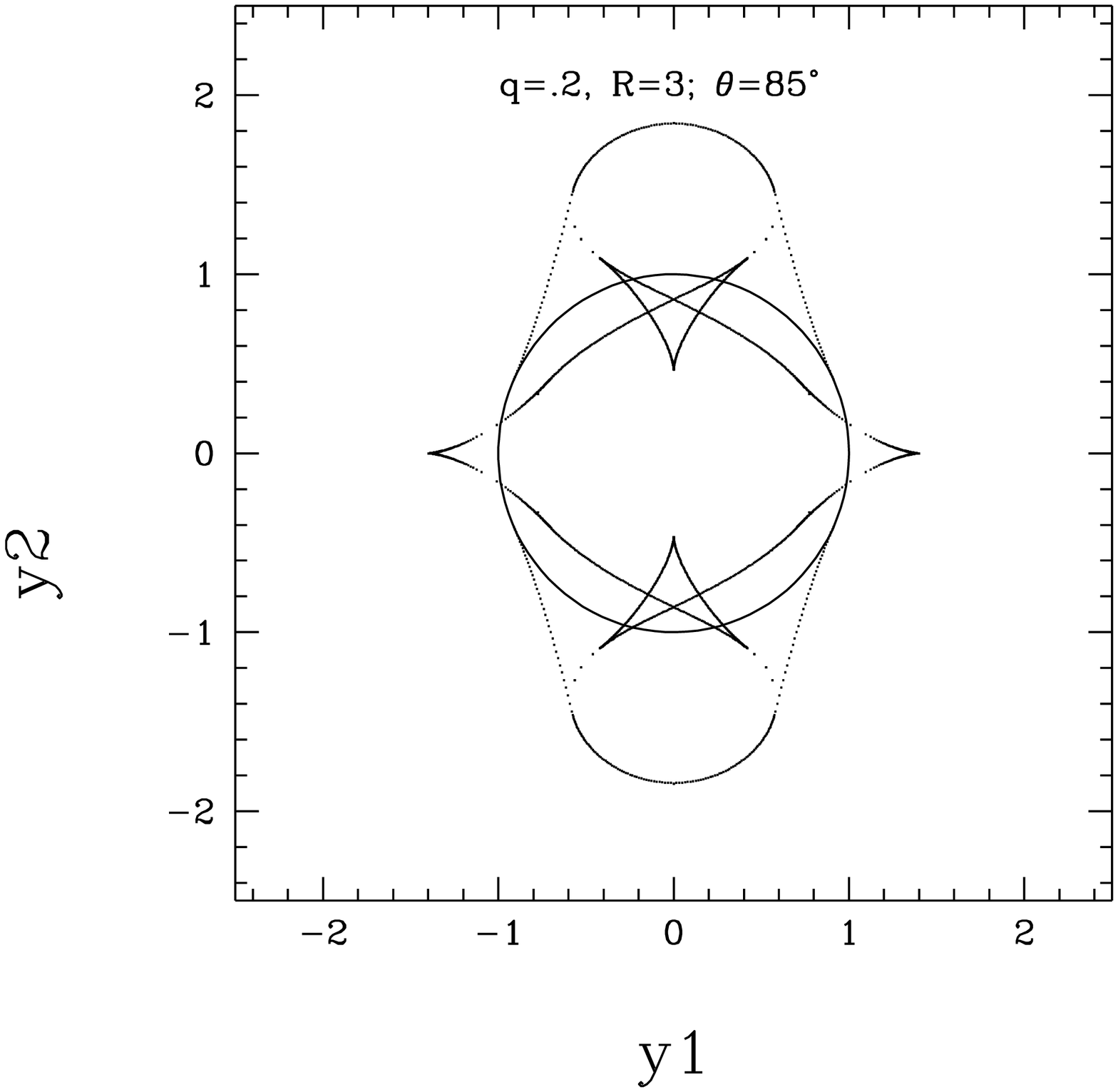}
\figcaption{$q=0.2$, $R=3$, 
$\theta=85^{\circ}> \theta_2=72.168^{\circ}$. 
The unit radius circles are the same as in Fig.1.
(a) Critical curve. (b) Caustic. }

\plottwo{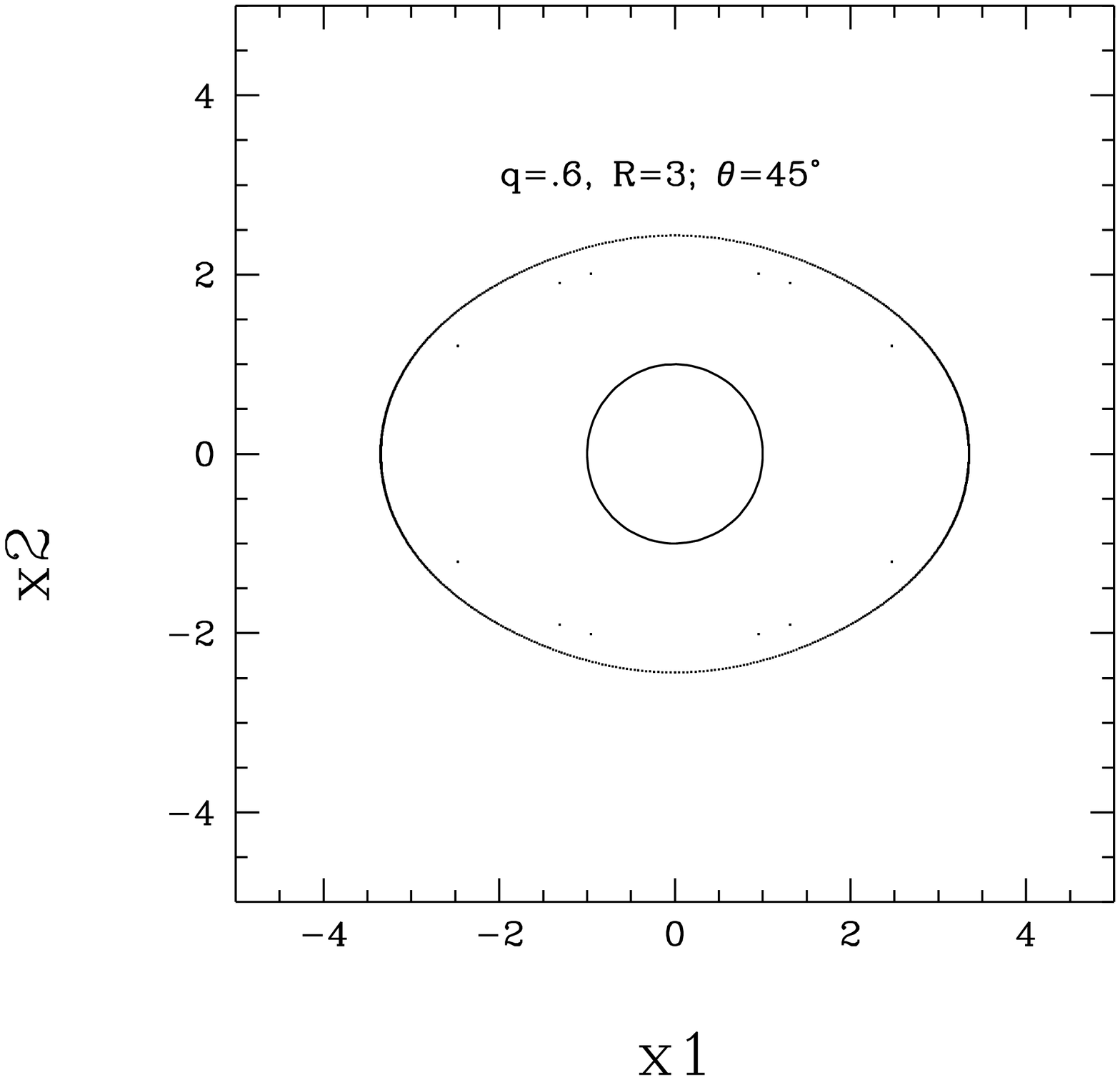}{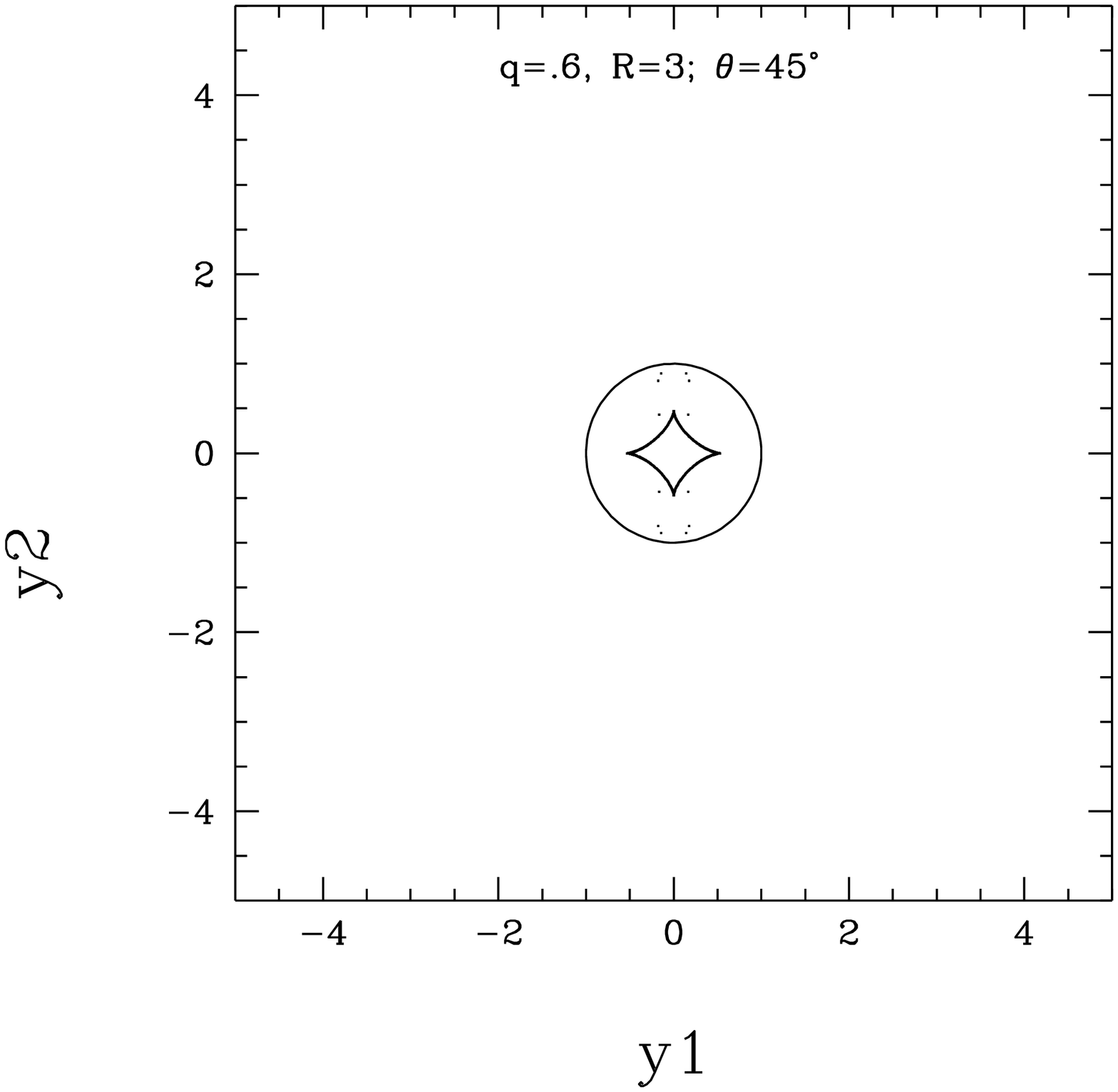}
\figcaption{$q=0.6$, $R=3$, 
$\theta=45^{\circ}< \theta_2=45.238^{\circ}$. 
The unit radius circles are the same as in Fig.1.
(a) Critical curve. (b) Caustic. }

\plottwo{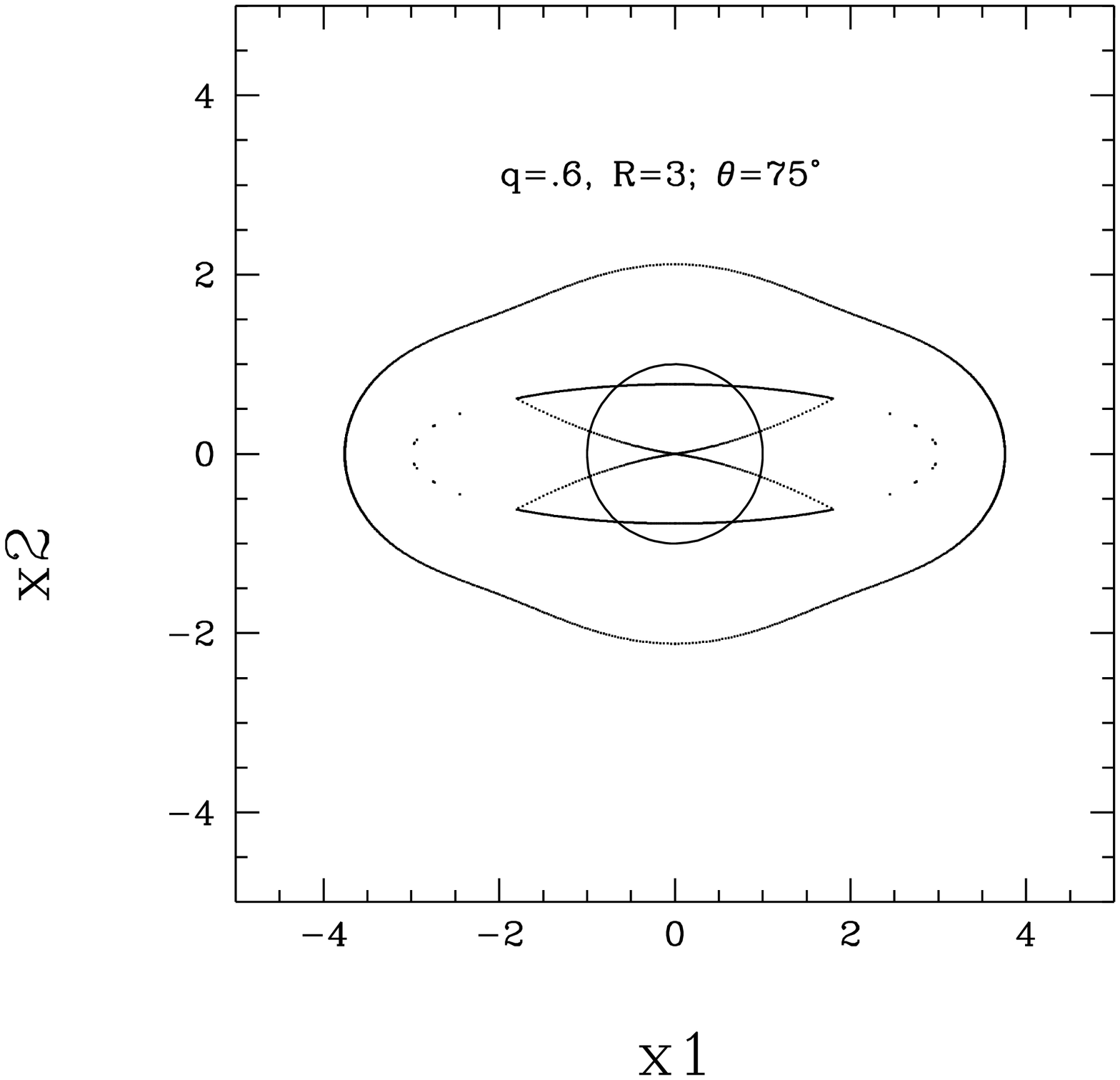}{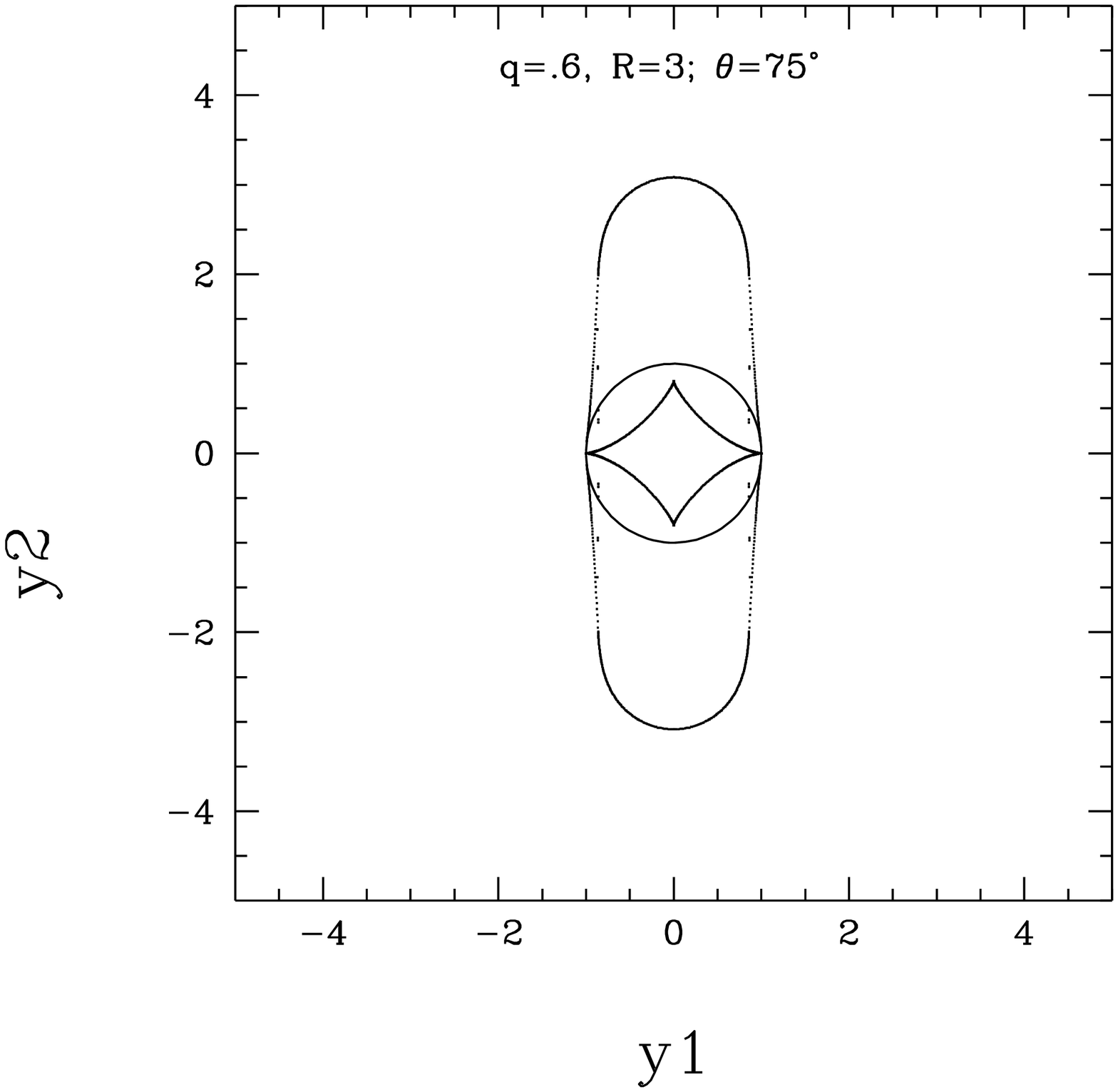}
\figcaption{$q=0.6$, $R=3$, 
$\theta_2=45.238^{\circ}< \theta=75^{\circ}< \theta_1=78.463^{\circ}$. 
The unit radius circles are the same as in Fig.1.
(a) Critical curve. (b) Caustic. }

\plottwo{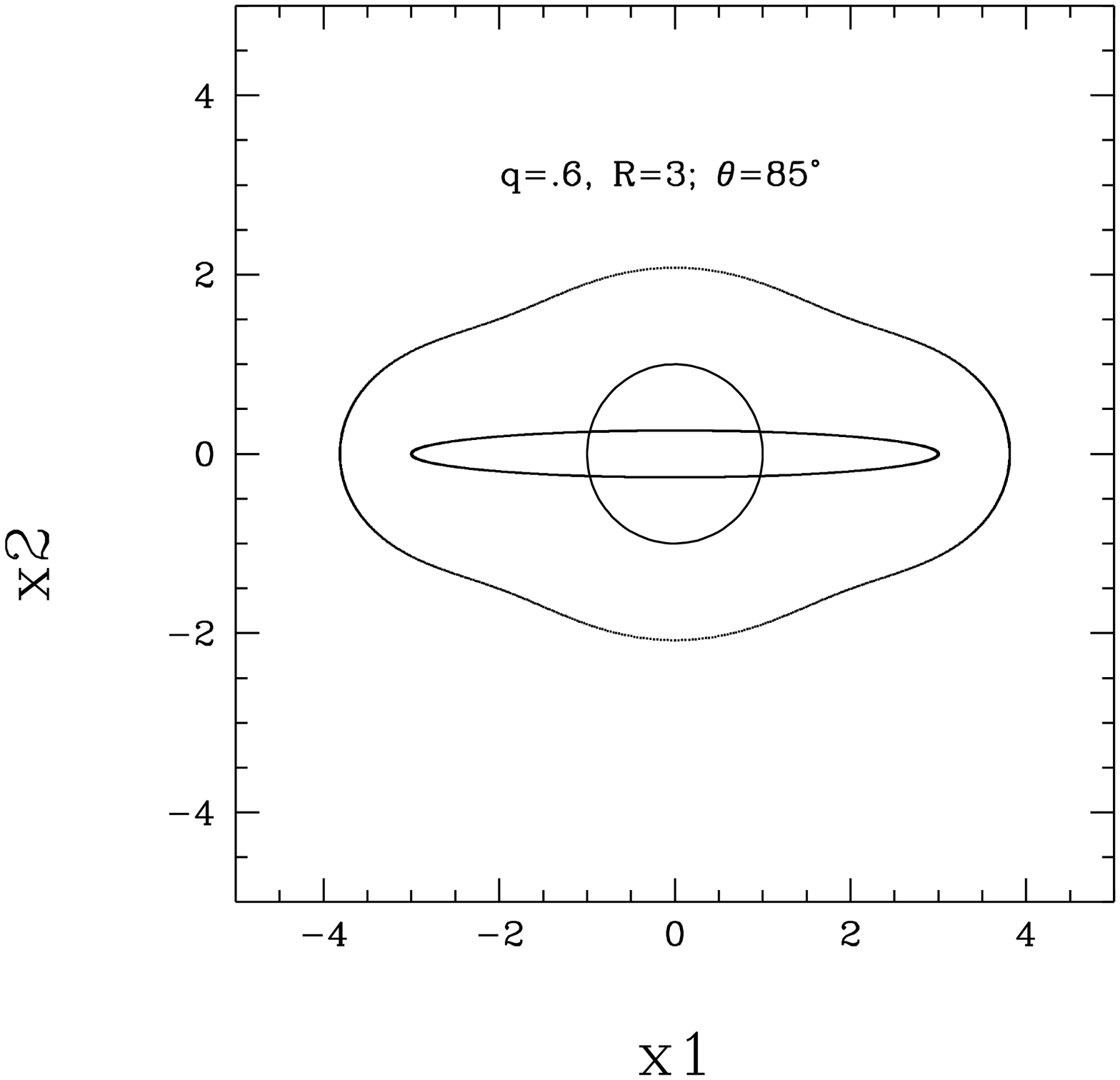}{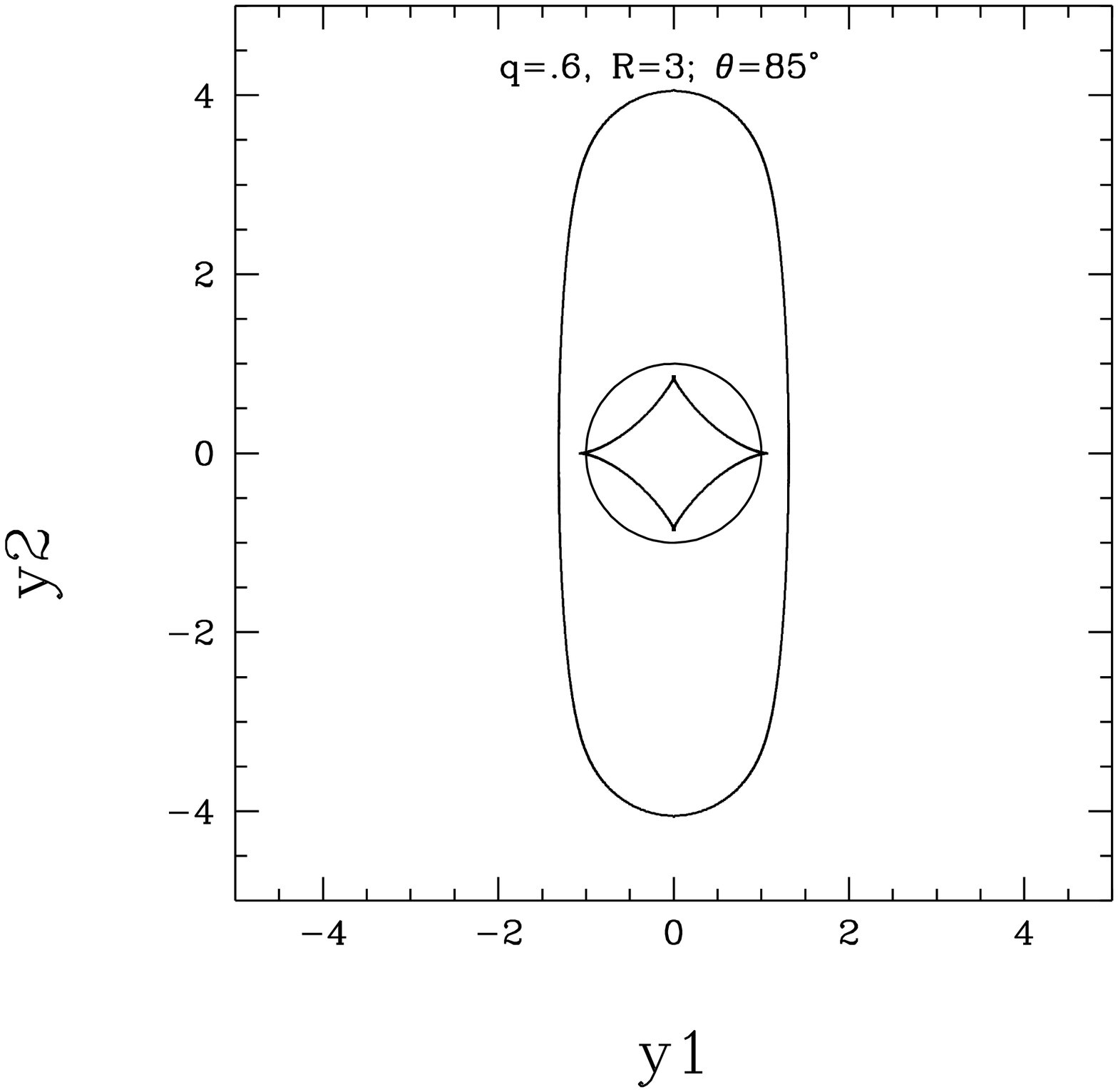}
\figcaption{$q=0.6$, $R=3$, 
$\theta=85^{\circ}> \theta_1=78.463^{\circ}$. 
The unit radius circles are the same as in Fig.1.
(a) Critical curve. (b) Caustic. }

\plotone{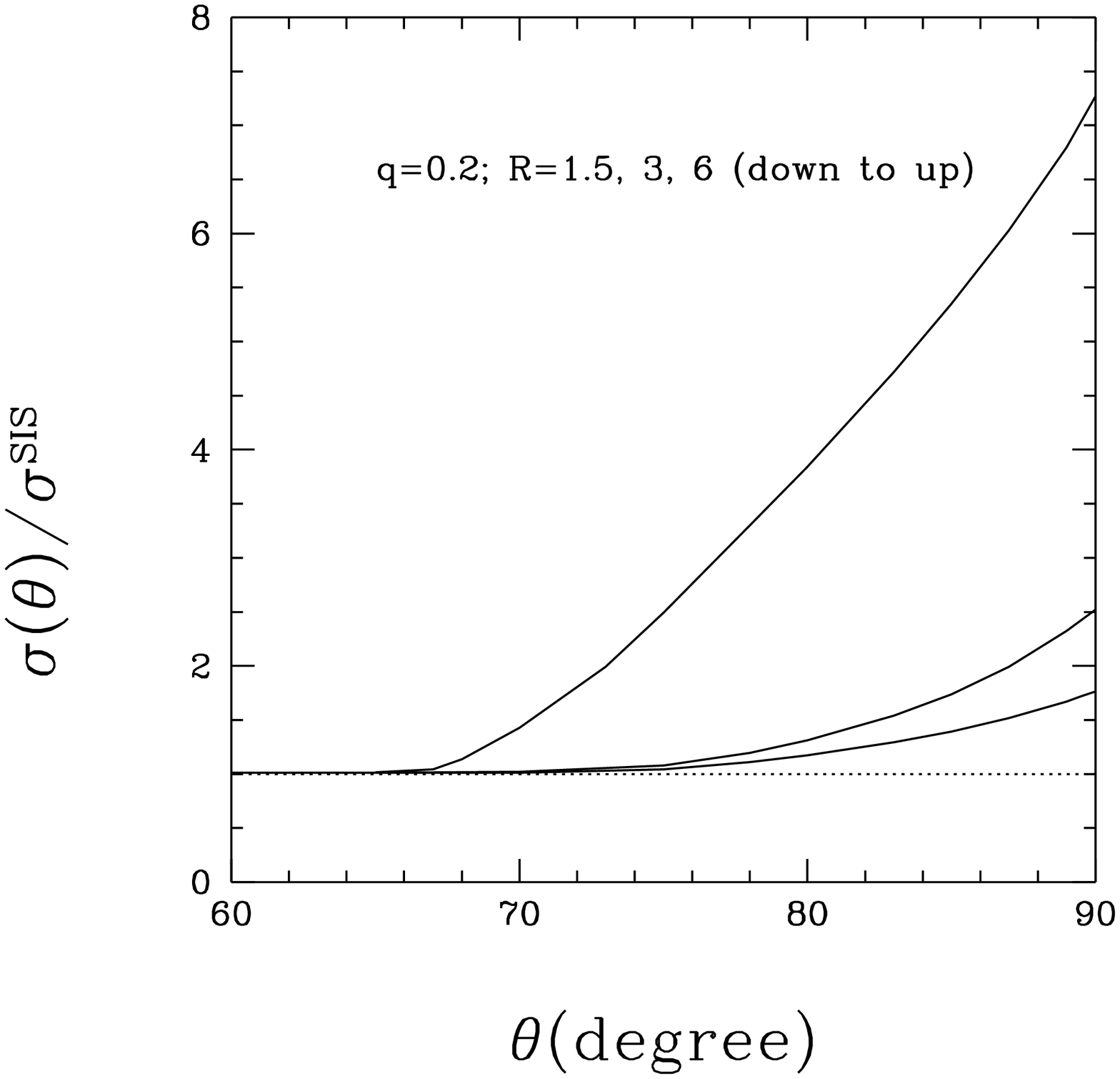}
\figcaption{The ratio of the multiple-image cross-sections 
for SIS plus inclined uniform disk and for SIS only, 
$\sigma(\theta)/\sigma^{SIS}$,
as a function of the inclination angle $\theta$, for $q=0.2$,
$R=1.5$, 3, 6.}

\plotone{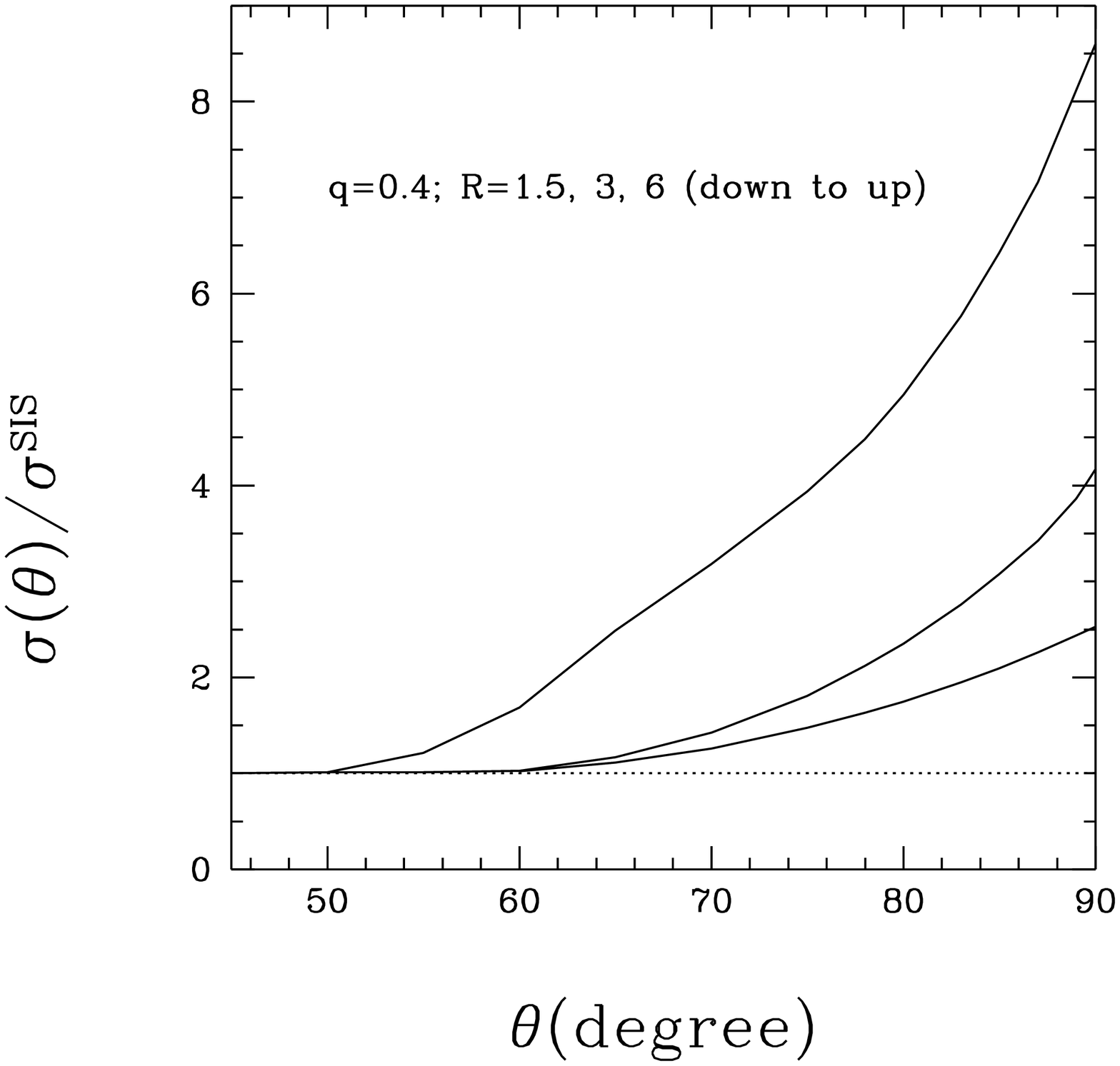}
\figcaption{Same as Fig.7, for $q=0.4$, $R=1.5$, 3, 6.}

\plotone{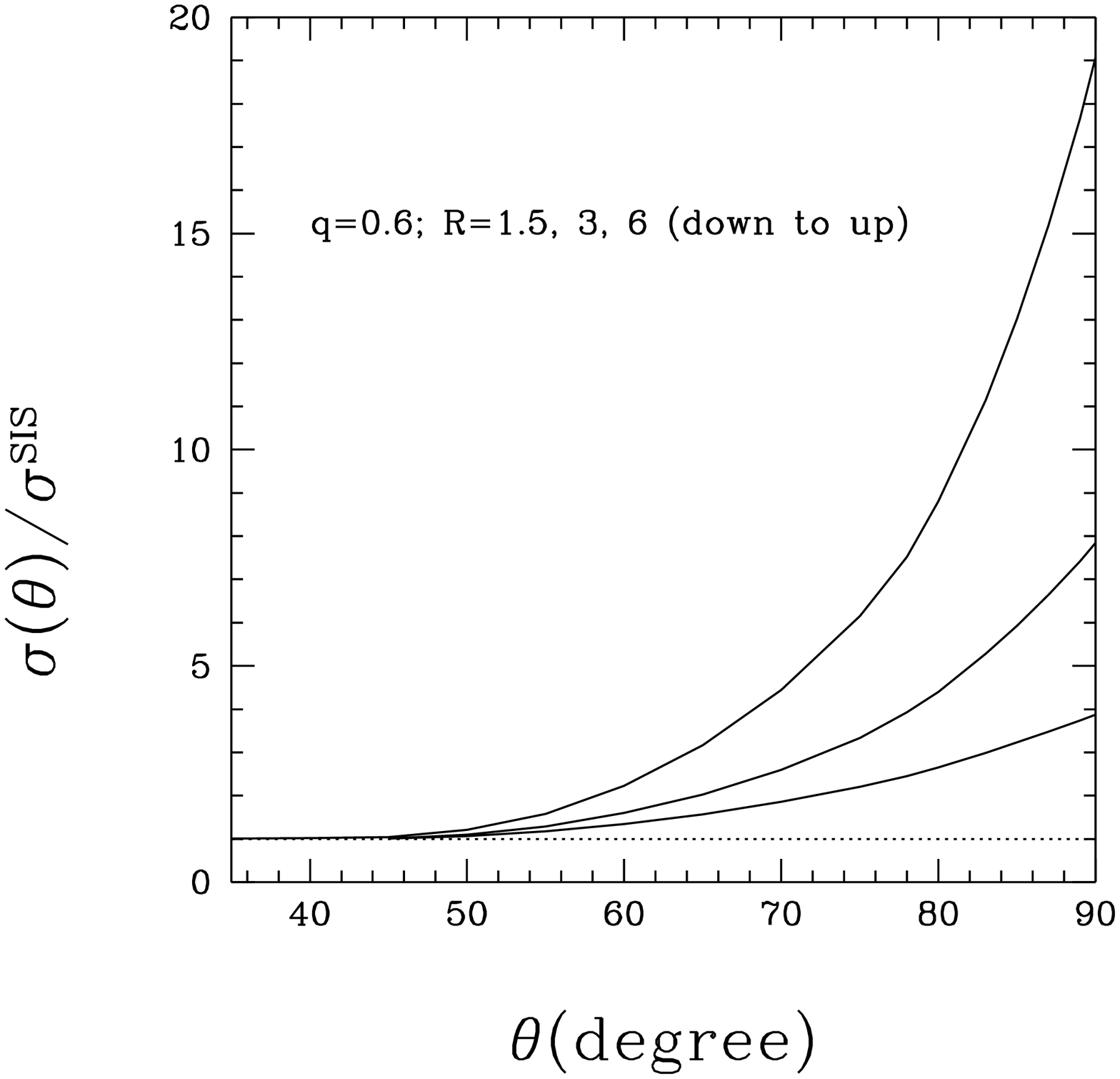}
\figcaption{Same as Fig.7, for $q=0.6$, $R=1.5$, 3, 6.}

\end{document}